# O/IR Polarimetry for the 2010 Decade (PSF): Science at the Edge, Sharp Tools for All

A Science White Paper for the:
**Planetary Systems and Star Formation (PSF) Science Frontiers Panel**
of the Astro2010 Decadal Survey Committee

Lead Authors:

Dan Clemens
Astronomy Department
725 Commonwealth Ave
Boston University
Boston, MA 02215
(617) 353 – 6140 (ph)
clemens@bu.edu

B-G Andersson
Stratospheric Observatory
  for Infrared Astronomy
NASA Ames Res. Center
Mail Stop 211-3
Moffett Field, CA 94035
(650) 604 6221 (ph)
bg@sofia.usra.edu

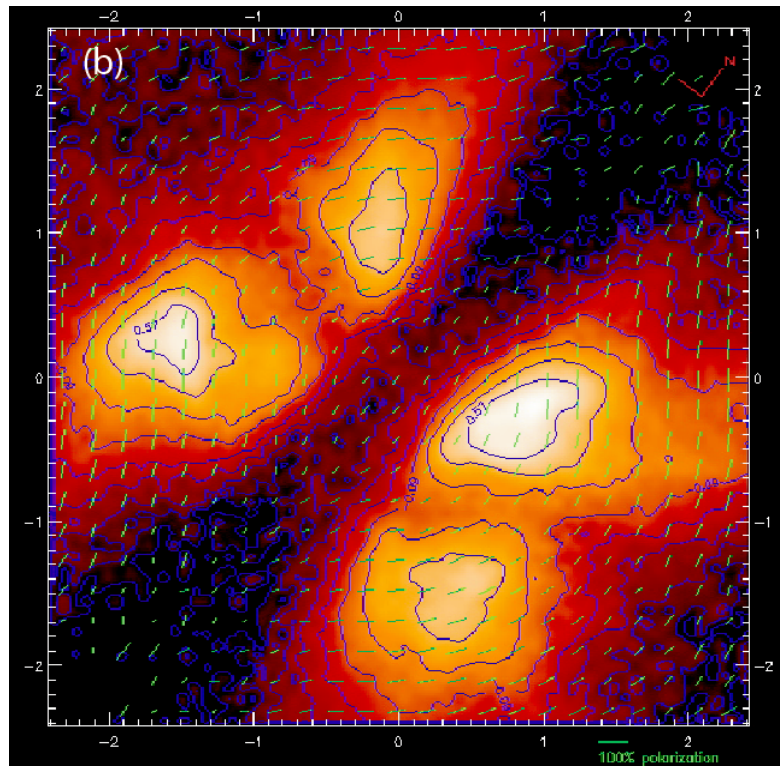

*NICMOS polarimetry of IRAS 04302+2247, a young Class I object embedded in L1536 (Cotera, et al. 2004; Hines et al. 2009, in prep.)*

*Contributors and Signatories*

| | |
|---|---|
| Andy Adamson | UKIRT, JAC, Hilo |
| David Axon | Rochester Institute of Technology |
| James De Buizer | SOFIA, NASA Ames |
| Alberto Cellino | Osservatorio Astronomico di Torino, Italy |
| Dean C. Hines | Space Science Institute, Corrales, NM |
| Jennifer L. Hoffman | University of Denver |
| Terry Jay Jones | University of Minnesota |
| Alexander Lazarian | University of Wisconsin |
| Antonio Mario Magalhaes | University of Sao Paulo, Brazil |
| Joseph Masiero | University of Hawaii |
| Chris Packham | University of Florida |
| Marshall Perrin | UCLA |
| Claudia Vilega Rodrigues | Inst. Nac. De Pesquisas Espaciais, Brazil |
| Hiroko Shinnaga | CalTech |
| William Sparks | STScI |
| John Vaillancourt | CalTech |
| Doug Whittet | RPI |

*Overview and Context: Polarimetry as a cross-cutting enterprise*

Photometry, spectroscopy, and polarimetry together comprise the basic toolbox astronomers use to discover the nature of the universe. Polarimetry established the Unified Model of AGN and continues to yield unique and powerful insight into complex phenomena. Polarimetry reveals the elusive magnetic field in the Milky Way and external galaxies, allows mapping of features of unresolved stars and supernovae, uncovers nearby exoplanets and faint circumstellar disks, and probes acoustic oscillations of the early universe.

Polarimetry is practiced across the full range of accessible wavelengths, from long wavelength radio through gamma rays, to provide windows into phenomena not open to photometry, spectroscopy, or their time-resolved variants. At some wavelengths, the U.S. leads the world in polarimetric capabilities and investigations, including ground-based radio, through the VLA and VLBA. At other wavelengths, the U.S. is currently competitive: in submm the CSO and the JCMT have historically pursued similar science problems.

In ground-based O/IR, the situation is considerably worse, with no optical or NIR polarimeters available on Gemini (Michelle is MIR, only) or any NOAO-accessed 4 m telescope, as the table below shows. Over the past decade and more, Canadian and European astronomers have enjoyed unique access to state-of-the-art polarimeters and have used this access to vault far past the U.S. in many science areas.

| Telescope | Aperture | Instrument | Waveband | Polar. Mode | U.S. Access ? |
|---|---|---|---|---|---|
| IRSF (SAAO) | 1.4m | SIRPOL | NIR | Imaging | No |
| Perkins (Lowell) | 1.8m | Mimir, PRISM | NIR, Optical | Imaging | Private |
| HST | 2.4m | WFPC2, ACS, NICMOS | Optical, Optical, NIR | Imaging | Yes |
| Nordic Optical | 2.5m | TURPOL | Optical | Photopol | No |
| MMT | 6.5m | MMTPOL | NIR | Imaging | Private |
| LBT | 2x8.4m | PEPSI | Optical | Spectropol | Private |
| Gemini | 8m | Michelle | MIR | Imaging | Yes |
| Keck | 10m | LRIS | Optical | Spectropol | Private |
| GTC | 10m | CanariCam | MIR | Imaging | No |

In space, NICMOS, ACS, and WFPC-2[1] on HST have permitted imaging polarimetry at modest precision, and may represent the most general purpose O/IR access for U.S. astronomers. Neither the Spitzer Space Telescope nor JWST provides, or will provide, any polarimetric capability.

The dwindling U.S. access to this crucial third leg of the light analysis tripod has also become self-fulfilling, as students receive little exposure to polarimetric techniques and scientific advances as the number of practitioners able to teach students declines.

Nevertheless, polarimetric studies in O/IR have already revealed a great deal about star and planet formation processes, stars and their evolution, the structure of the Milky Way, and the nature and origin of galaxies and their active nuclei – details which cannot be discovered using pure photometric or spectroscopic methods. For example, NIR imaging polarimetric studies by the SIRPOL group (e.g., Tamura et al. 2007) reveal the details of the magnetic fields lacing nearby, star-forming molecular cloud cores and the embedded reflection nebulae and disks associated with their newly formed stars. Further, the race to find and image exoplanets will use polarimetry internal to the two extreme adaptive optics coronagraphs now under construction: SPHERE/ZIMPOL (for the VLT: Joos 2007) and GPI (for Gemini: Macintosh et al. 2006).

---

[1] NICMOS and much of ACS are currently off-line until Servicing Mission 4 (SM4); WFPC-3 will replace WFPC-2 but will have no polarimetric capability.

Theoretical efforts have recently advanced our understanding of the origin of dust grain alignment by demolishing old theory and offering new, testable predictions. The new paradigm of radiative aligned torques (e.g., Lazarian & Hoang 2007; Hoang & Lazarian 2008) is removing old doubts concerning magnetic alignment while leaving a wide parameter space open for observational testing and future theoretical refinement.

Polarimetric modeling has also entered a modern age, one characterized by huge model grids, spectacular dynamic ranges, and closer coupling to observational measurements. The active debate on the relative importance of magnetohydrodynamics (e.g., Li et al. 2004) vs. pure hydrodynamics (e.g., Padoan & Nordlund 2002) is shining new light into the nature of magnetic fields in the ISM and in star formation. Meanwhile, radiative transfer modeling guides detailed interpretation of complex polarized line profiles from aspherical stellar winds (e.g., Harries 2000) and supernovae (e.g., Hoeflich 2005; Kasen et al. 2006; Hoffman 2007). This upcoming decade is sure to see careful, detailed comparisons between high-resolution model simulations and observational data that will lead to sharp new insights into many astrophysical scenarios.

The promise evident in the new, niche polarimetric instruments and the surveys they will perform will drive cutting-edge science in the upcoming decade. Yet finding answers for many key questions requires open community access to general purpose, precision polarimeters on large telescopes, as well as opportunities for student training.

*Example Polarimetry Science Areas for the next decade*

Within the broad area covered by Planetary Systems and Star Formation, polarimetric studies for the upcoming decade will focus on answering some key questions: (1) How do magnetic fields affect star formation, in particular, accretion onto protostars and development of disks and jets? (2) Do massive stars form predominantly via accretion disks that require attendant outflows, or via mergers of lower mass cores or protostars? (3) What processes grow ISM-type dust grains to be larger grains in disks and from there to planetesimals? (4) What can polarimetry reveal about life's origins from measurements of chirality? (5) How do we accurately assess the danger to this planet from asteroids in our solar system? Here, we outline four example areas ripe for advance in the upcoming decade using polarimetric techniques.

- Measure magnetic fields within dense cores and near forming YSOs.

Star formation theories often ignore magnetic fields, yet the fields may be key regulators of infall for forming stars. Magnetic fields in regions of star formation are traced through the polarization emitted by dust grains aligned with magnetic fields, with different wavelengths probing different depths into star forming regions. Sub-mm to radio wavelengths yield only part of the picture, as they trace cool dust far from new YSOs. Yet, light with wavelengths shorter than 5 μm cannot easily escape from these deeply embedded regions. The magnetic field information in the inner regions of massive cores is therefore missing. Polarization measurements spanning all wavelengths for which dust emits radiation are crucial. New MIR (e.g., the Michelle instrument at Gemini North; Packham et al. 2007) and FIR polarization measurements are needed to complement those to be obtained at longer ground-based wavelengths (i.e., SMA, ALMA). Such a panchromatic set of polarimetric observations will reveal the first really 3-D picture of the distribution of, and roles played by, magnetic fields in star forming regions. Recently, probes of low-mass star forming regions that utilize this multi-wavelength approach have been performed (e.g., NGC 2024: Kandori et al. 2007), showing the efficacy of the approach. In the upcoming decade, extensive new polarization observations in the 20-100 μm wavelength regime (from a combination of ground-based and SOFIA-based instruments) would allow testing detailed theories of the role of magnetic fields for both low- and high-mass star formation.

- Reveal structures and compositions of YSO, protoplanetary, and TTauri disks

Dust disks around newly forming protostars, evolving YSOs, and more revealed TTauri stars are highly polarized in scattered light. Thus, polarimetry provides diagnostics of disk geometry (inclination, line of nodes, near-side orientation) and physical properties of the scattering particles. For edge-on disks, polarimetry breaks degeneracies, enabling precision measurement of particle size and porosity (e.g., Graham et al. 2007). Key goals for the upcoming decade will focus on ascertaining the detailed structure of these disks, for example whether they are flared for lower mass stars, what their magnetic configurations are and the role of the field in the disks, how and where winds and outflows are launched from the disks, and how dust grains in disks evolve with time.

Planets form here, so assessing disk structures and dynamics for clues to planet formation and migration processes will be vital. Polarimetry provides direct information about the composition of dust in the belts in protoplanetary disks, revealing stratification of material during planet building phases.

At MIR/FIR wavelengths, disk opacities are moderate and polarization from the disk surfaces can be measured, in particular at the locations where outflows are likely launched. MIR spectropolarimetry is needed to assess compositions of disk surface material. For example, the flatness of the 10 μm silicate feature depends on particle size and reveals particle growth throughout the disk. Likewise, polarimetry of the 20 μm silicate feature and PAH features (Sironi & Draine 2009) in the MIR/FIR can uniquely probe disk chemistry.

For high-mass star formation, MIR polarimetry reveals the inner structures of embedded massive YSOs. But, NIR polarimetry probes light scattered by grains in outflow cones and larger envelopes, enabling tests for multiple light sources and disentangling scattered (thus polarized) light emission from unpolarized emission.

- Astrochirality and the Origin of Life.

Interstellar dust holds the elements needed to make planets and life. Might the dust grains also serve as key production sites for life's biomolecules? A chiral molecule possesses two configurations (L and D) that are chemically equivalent but of opposite handedness and so interact differently with light. Terrestrial biology is homochiral, whereas non-biological organic matter has no chiral preference. Detection of a preference for the L-form in meteoritic amino acids (Pizzarello 2004) means homochirality may arise from the early star-formation environment of the Solar System. Amino acids synthesized in lab experiments using UV irradiation of interstellar ices (Meierhenrich & Thiemann 2004) show that circularly polarized incident radiation induces chiral asymmetry.

O/IR imaging polarimetry (both circular and linear) of star formation regions in the upcoming decade is urgently needed to test this connection in astrophysical settings. The key question to answer is this: Does the presence of predominantly circularly polarized light impinging on ice-covered dust grains set the basis of homochirality and its influence on biogenesis? Most observations to date have focused on just one target - the Orion Nebula, where very high levels of polarization are observed (Bailey et al. 1998; Fukue et al. 2009, and references therein). A wider selection of astrophysical laboratories needs to be probed for astrochirality in the upcoming decade to inform our understanding of life's origins.

- <u>Assessing "Death from the Skies" – pinning down asteroid albedos and sizes</u>.
  Despite several surveys to inventory the population of near-Earth objects (NEOs) and asteroids as potential harmful objects (e.g., Spacewatch: Gehrels & Jedicke 1996; LONEOS: Howell et al. 1996), the actual diameters and masses of the detected objects remain unknown. Polarimetric observations of asteroids represent a powerful technique to determine the albedos, and hence the sizes of these objects (e.g., Morbidelli et al. 2002; Delbo et al. 2007), to better assess their kinetic energy, and to rate the danger posed by a given object based on its Earth impact probability and delivered energy (for the importance of albedo in assessing risk, c.f., Napier et al. 2004). A dramatic increase in the number of measured asteroid polarization-phase curves (i.e., via an LSST polarimetric campaign[2]) is needed to clarify the population divisions among the Main Belt and NEOs (e.g., Stuart & Binzel 2004). While qualitatively well established, the relation between polarimetric observables and asteroid albedos needs to be better quantified and calibrated. The multi-wavelength polarimetric surveys needed to assess the asteroid properties will also open new fields of investigation for other minor bodies, including comets and their dust evolution, in the solar system.

*Exceptional Discovery Potential Area: Exoplanets & their atmospheres*

There is one area which offers the potential for spectacular advancement using polarimetric techniques - that of exoplanets, in particular revealing the details of the structures and contents of their atmospheres. Polarimetry is already being employed to boost the planet-star contrast ratio to aid detection, but polarimetry is also an important way to probe exoplanets atmospheres. Though the atmospheres of eclipsing exoplanets have been detected and are being studied spectroscopically, polarimetry (e.g., Stam et al. 2004) offers the best access to the vastly larger non-eclipsing sample of exoplanets that are too close to their stars to be directly imaged. Variations in phase, as well as surface and atmosphere compositions and structures, all have specific polarization signatures for solar system planets (Hansen & Hovenier 1974), Hot Jupiters (Seager et al. 2000), and have recently been modeled for Earth-like exoplanets (Stam 2008). The polarized light wavelength dependence also reveals atmospheric scattering processes.

Exoplanet atmospheres are being studied from the ground using high-precision polarimetry (e.g., with PlanetPol; Hough et al. 2006) with application to "Hot Jupiters." The controversy regarding polarimetric detection of HD 189733b by Berdyugina et al. (2008) [see recent modeling by Sengupta (2008) and a very recent claim of non-detection by Wiktorowicz (2009)] and the upper limits for 55 Cnc and tau Boo reported by Lucas et al. (2008) underscore the potential of, and great interest for, polarimetry to address issues of exoplanet sizes, orbits, and atmospheres.

*What is Needed to Meet the Science Goals within the Decade*

What key observations, theory, and instrumentation are needed to achieve the science goals within the next decade? Our evaluation of the upcoming opportunities, challenges, and technical readiness leads to the following recommendations:

---

[2] See the All-Sky Polarimetry Survey White Paper (on http://astroweb.iag.usp.br/~mario/) submitted to the LSST Consortium

1. *Build precision polarimetric capability into new O/IR instruments for large telescopes and space missions. Design polarimetry in from the beginning, not as "add-ons".*
   - New "niche" instruments, such as GPI, SPHERE/ZIMPOL, and HiCIAO on Subaru rightly exploit polarimetry to meet their exoplanet and circumstellar disk objectives, but general purpose instruments with polarimetric capability are lacking at virtually all large, open-access US telescopes - ground-based, airborne, and space-based, especially in the infrared.
   - Key science questions cannot be answered unless US astronomers have access to precision (photon-noise limited) polarimetric capability. To retain precision capability, polarimetric capabilities must be a considered in the initial design of the instruments, not as a later "add-on".
   - Exoplanet characterizations may favor bluer wavelengths, to enhance detection of Rayleigh scattering, but fractional polarizations are low, requiring access to the largest telescopes to enable sampling adequate volumes.

2. *Encourage polarimetric surveys with LSST.*
   - So much of the sky has never been explored polarimetrically that it represents a "new frontier" for optical wavelengths. Many key science advances are in the Polarimetry White Paper to the LSST Consortium[2].
   - Asteroid albedo determinations will be best performed using wide-field synoptic survey on and off the ecliptic.
   - LSST surveys offer one of the best chances of getting polarimetric data into the hands of the largest number of researchers and students.

3. *Develop polarimetric O/IR synoptic and survey capabilities on intermediate-size telescopes to study YSOs/disks, probe their time evolution, and to promote student training in instrumentation and polarimetric observations.*
   - To be able to compete scientifically, we must invest in the next generation of young astronomers who will use polarimetry as a powerful tool in their light analysis toolbox and who will understand polarimetric light analysis well enough to guide future instrument development.
   - Wide-field polarimetric surveys, especially in the infrared, are needed to probe into extincted regions of star formation to discover and measure circumstellar and protostellar environments.
   - Synoptic polarimetric observational data sets are crucial to understanding phenomena with complex geometry and/or time evolution.
   - Obtaining ground-based calibrating polarimetric observations are crucial to the calibration of existing and future space-based polarimeters.

*Final Thought*

The U.S. astronomical community has lost opportunities to advance key science areas as a result of down-selects of instrument capabilities or lack of will to commission polarimetric modes on instruments. The investment is minor, the expertise is available in the community, and the rewards are tangible. We are excited by the recent momentum favoring polarimetric studies and capabilities and believe the upcoming decade will see the various polarimetric techniques together become a strong, necessary component of astronomers' light analysis toolbox.

*Bibliography and References*

Bailey, J., Chrysostomou, A., Hough, J. H., Gledhill, T. M., McCall, A., Clark, S., Menard, F., & Tamura, M. 1998, "Circular Polarization in Star-Formation Regions: Implications for Biomolecular Homochirality," Science, 281, 672

Berdyugina, S. V., et al. 2008, "First Detection of Polarized Scattered Light from an Exoplanetary Atmosphere," ApJ, 673, L83-L86

Delbò, M., Cellino, A., Tedesco, E. F. 2007, "Albedo and size determination of potentially hazardous asteroids: (99942) Apophis," Icarus, 188, 266

Fukue, T., et al 2009, "Near-Infrared Circular Polarimetry and Correlation Diagrams in the Orion Becklin-Neugebauer/Kleinman-Low Region: Contribution of Dichroic Extinction," ApJ, 692, L88

Gehrels, T., & Jedicke, R. 1996, "The Population of Near-Earth Objects Discovered by Spacewatch," Earth, Moon and Planets, 72, 233-242.

Graham, J. R., Kalas, P. G., & Matthews, B. C. 2007, "The Signature of Primordial Grain Growth in the Polarized Light of the AU Microscopii Debris Disk," ApJ, 654, 595

Hansen, J. E., & Hovenier, J. W. 1974, "Interpretation of the polarization of Venus," J. Atmos. Sci., 31, 1137

Harries, T. J. 2000, "Synthetic line profiles of rotationally distorted hot-star winds," MNRAS, 315, 722

Hoang, T., & Lazarian, A. 2008, "Radiative torque alignment: essential physical processes," MNRAS, 388, 117

Hoeflich, P. 2005, "Radiation hydrodynamics in supernovae," Ap&SS, 298, 87

Hoffman, J. L. 2007, "Supernova polarization and the Type IIn classification," in AIP Conf. Ser. 937 (Supernova 1987A: 20 Years After: Supernovae and Gamma-Ray Bursters), eds. S. Immler, K.W. Weiler, & R. McCray (New York: AIP), 365

Hough, J.H., et al. 2006, "PlanetPol: A Very High Sensitivity Polarimeter", PASP, 118, 1302-1318.

Howell, S. B., Koehn, B., Bowell, E., & Hoffman, M. 1996, "Detection and Measurement of Poorly Sampled Point Sources Imaged With 2-D Array," AJ, 112, 1302.

Packham, C., & Jones, T. J. 2008, "MMT-Pol: an adaptive optics optimized 1-5μm polarimeter," SPIE, 7014, 180

Joos, F. 2007, "Polarimetric direct detection of extra-solar planets with SPHERE/ZIMPOL," Proceedings of the conference "In the Spirit of Bernard Lyot: The Direct Detection of Planets and Circumstellar Disks in the 21st Century," lyot.confE, 28

Kandori, R., Tamura, M., Kusakabe, N., Nakajima, Y., Nagayama, T., Nagashima, C., Hashimoto, J., Ishihara, A., Nagata, T., & Hough, J. H. 2007, "Near-Infrared Imaging Polarimetry of the Star-Forming Region NGC 2024," PASJ, 59, 487-506

Kasen, D., Thomas, R., & Nugent, P. 2006, "Time dependent Monte Carlo radiative transfer calculations for 3-dimensional supernova spectra, lightcurves, and polarization," ApJ, 651, 366

Lazarian, A., & Hoang, T. 2007, "Radiative torques: analytical model and basic properties," MNRAS, 378, 910

Li, P. S., Norman, M. L., Mac Low, M.-M., & Heitsch, F. 2004, "The Formation of Self-Gravitating Cores in Turbulent Magnetized Clouds," ApJ, 605, 800

Lucas, P. W., et al. 2008, "Planetpol polarimetry of the exoplanet systems 55 Cnc and tau Boo," arXiv:0807.2568

Macintosh, B., et al. 2006, "The Gemini Planet Imager," SPIE, 6272, 18

Morbidelli, A., R. Jedicke, W.F. Bottke, P. Michel, E.F. Tedesco 2002, "From Magnitudes to Diameters: The Albedo Distribution of Near Earth Objects and the Earth Collision Hazard," Icarus, 158, 329-342.

Meierhenrich, U. J., & Thiemann, W. H.-P. 2004, "Photochemical Concepts on the Origin of Biomolecular Asymmetry," OLEB, 34, 111

Napier, W. M., Wickramasinghe, J. T., & Wickramasinghe, N. C. 2004, "Extreme albedo comets and the impact hazard," MNRAS, 355, 191-195

Packham, C., et al. 2007, "Gemini mid-IR polarimetry of NGC 1068: polarized structures around the nucleus," ApJ, 661, L29

Padoan, P., & Nordlund, Å. 2002, "The Stellar Initial Mass Function from Turbulent Fragmentation," ApJ, 576, 870

Pizzarello, S. 2004, "Chemical Evolution and Meteorites: An Update," Orig. Life & Evol. Biosph., 34, 25-34

Seager, S., Whitney, B. A., & Sasselov, D. D. 2000, "Photometric Light Curves and Polarization of Close-in Extrasolar Giant Planets," ApJ, 540, 504-520

Sengupta, S. 2008, "Cloudy Atmosphere of the Extrasolar Planet HD 189733b: A Possible Explanation of the Detected B-Band Polarization," ApJ, 683, L195-198.

Sironi, L., & Draine, B. T. 2009, "Polarized Infrared Emission by Polycyclic Aromatic Hydrocarbons resulting from Anisotropic Illumination," arXiv0901.4558

Stam, D. M., Hovenier, J.W., and Waters, L. B. F. M. 2004, "Using polarimetry to detect and characterize Jupiter-like extrasolar planets," A&A, 428, 663–672

Stam, D. M. 2008, "Spectropolarimetric signatures of Earth-like extrasolar planets," A&A, 482, 989-1007

Stuart, J. S., & Binzel, R. P. 2004, "Bias-corrected population, size distribution, and impact hazard for the near-Earth objects," Icarus, 170, 295-311

Tamura, M., et al. 2007, "Near-Infrared Imaging Polarimetry of the NGC 2071 Star-Forming Region with SIRPOL," PASJ, 59, 467

Wiktorowicz, S. J. 2009, "Non-Detection of Polarized, Scattered Light from the HD 189733b Hot Jupiter," arXiv0902.0624